\documentclass[usenatbib]{basi}

\usepackage[british]{babel}
\usepackage[varg]{txfonts}
\usepackage{rotating}
\usepackage{dcolumn}

\begin{document}
\title[Infrared studies of the intermediate polar WX Pyx]{A study of the near-infrared modulation at
spin and orbital periods in the intermediate polar WX Pyx}
\author[V. H. Joshi et. al.]%
       {V. H. Joshi\thanks{email: \texttt{vjoshi@prl.res.in}},
       N. M. Ashok and D. P. K. Banerjee\\
       Physical Research Laboratory, Ahmedabad 380 009, India}

\pubyear{2011}
\volume{00}
\pagerange{\pageref{firstpage}--\pageref{lastpage}}

\date{Received \today}

\maketitle

\label{firstpage}

\begin{abstract}
We present near-infrared $J$  band photometric observations of the intermediate polar WX Pyx. The
frequency analysis indicates the presence of a period at 1559.2 $\pm$ 0.2 seconds which is attributed to
the spin period of the white dwarf. The spin period inferred from the infrared data closely
matches that determined from X-ray and optical observations. WX Pyx is a system whose orbital
period has not been measured directly and which is not too well constrained. From the IR
observations, a  likely peak at 5.30 $\pm$ 0.02 hour is seen  in the power spectrum of the object. It is
suggested that this corresponds to the orbital period of the system. In case this is indeed the true orbital period, some of the system physical parameters may be estimated. Our analysis indicates that
the  secondary star is of M2 spectral type and the distance to the object is 1.53 kpc. An upper limit of  30$^\circ$ for the angle of inclination of
the system is suggested. The mass transfer rate  and the magnetic moment of the white dwarf
are estimated to be (0.95 - 1.6)$\times10^{-9} M_\odot$ yr$^{-1}$ and (1.9 - 2.4)$ \times 10^{33}$ G cm$^{3}$ respectively.
\end{abstract}


\section{Introduction}\label{s:intro}
Intermediate polars (IPs) are a sub-class of magnetic cataclysmic variable stars
(mCVs). An IP  comprises of a  semi-detached binary system  consisting of a  white dwarf primary
star and a late-type Roche-lobe filled secondary star; the white dwarf is believed to have  a
moderate magnetic field strength of  B $\sim$ 1 - 10 MG \citep{Patterson94}. The rotation period of
the white dwarf is not synchronized with the binary orbital period. The mass transferred from the
secondary to the primary, by the process of  Roche lobe overflow, eventually accretes onto or
near the magnetic poles of the white dwarf through the magnetic field lines. The shock heated
plasma produces X-rays by the cooling of electrons near the white dwarf's surface through
free-free interactions \citep{Evans07}. Recently, \citet{Bednarek10} have proposed
that $\gamma$ rays could also be produced when the accelerated hadrons are convected to the white
dwarf surface and interact with the dense matter. A comprehensive review of IPs may be found in
\citet{Warner95}.

WX Pyx was identified as 1E 0830.9-2238 in the Galactic plane survey of Einstein X-ray observatory by \citet{Hertz84}. The blue excess seen in the object and also the nature of the optical spectra, which displayed  emission lines including Hydrogen Balmer lines,  He{\sc i}   and He{\sc ii} - lines which are typically seen in the spectra of CVs -  suggested the object as a CV candidate (Hertz et al. 1990). The strong presence of the high-excitation  He{\sc ii} 4686 \AA line in the optical spectrum also  suggested  the possibility of a magnetic nature for the system \citep{Hertz90}. From an analysis of the  optical light curve \citet{ODonoghue96} found a stable period of $\sim$ 26 minutes which was attributed  to the spin period of the white dwarf and confirmed the IP nature of the WX Pyx system. They suggested a value of of $\sim$ 6 to 9 hours for the orbital period, based on spectroscopic and photometric data, but stressed that this result was not robust and should be viewed with caution.  Though there are no direct X-ray observations of WX Pyx available
till date, the object was accidentally detected 10.5 arc-minutes off-axis during a pointed observation of NGC
2613 by XMM-Newton. \citet{Schlegel05} analyzed data related to this serendipitous X-ray detection and found a
spin period of 1557.3 s which matched well with the optical spin-period estimate of \citet{ODonoghue96}. An orbital period of $\sim$5.54 hours was inferred indirectly from the separation of spin-orbital side band frequencies in the power spectrum.

The present study is partially motivated by the fact that there are no infrared studies of WX Pyx. More importantly, we had hoped to determine more robustly the orbital period of the system for which no direct observational evidence is available. Such a determination would allow  many of the system parameters and mode of accretion of WX Pyx to be better estimated. In this study we present  photometric observations of  near-IR $J$ band and undertake a time series analysis to estimate the orbital and spin period of the object.

\section{Observations and data analysis}\label{s:obs}
Near infrared $J$ band photometry of WX Pyx was done using the Mt.Abu 1.2-m telescope.
The object was observed for a total duration of 22 hours  spanning  six nights from
December 2007 to February 2010. The observations were taken using the  Near-Infrared
Imager/Spectrometer which uses a 256x256 HgCdTe NICMOS3 array with a FOV of 2x2 arc-min$^2$.
The telescope was dithered at 5 different positions during observations, as is customary for IR observations (for e.g. \citet{Das08}), to produce median sky-cum-dark frames for individual images. Several selected field stars were always kept in the field of view while dithering to enable differential photometry to be done. Frames of smaller durations were co-added  when single frames of larger duration were not available. Flat fielding,
using twilight flats, was done to ensure a proper pixel response of the detector. An appropriate
median sky-cum-dark image was generated and subtracted from the object frames. Instrumental
magnitudes, obtained from these sky-subtracted frames using aperture photometry, were finally
flux calibrated by comparing with  several field  stars whose  lightcurves  had previously been
checked for photometric stability.

The routines used for data reduction are based on the  Interactive Data
Language (IDL) package. Specifically, APER procedure from GSFC IDL astronomy user's library was
used for synthetic aperture photometry. The log of the observations is given in Table 1,
which gives the date of observation, filter information, exposure time,
total observation coverage and number of spin cycles covered.

\begin{table}
 \centering
  \caption{Log of the photometric observations.}
  \medskip
  \begin{tabular}{@{}cccccc@{}}
  \hline
   Date of      & Reduced & Filter   & Exp.      & Time  & Spin  \\
   observation  & JD       &          & time     & cover- & cycles \\
   (dd/mm/yyyy) &         &          & (sec)    & age (hr)   & covered           \\
  \hline
    18/12/2007	& 54452   & J	       & 180	  & 4.5	     & 10         \\
    19/12/2007	& 54453   & J	       & 180      & 3.9	     & 9          \\
    18/01/2008	& 54483   & J	       & 60       & 3.3	     & 7.5        \\
    19/03/2009	& 54909   & J	       & 60       & 4.1	     & 9.5        \\
    19/03/2009	& 54910   & J	       & 60	      & 4.0	     & 9.5        \\
    22/02/2010	& 55250   & J	       & 180	  & 2.2	     & 5          \\
 \hline
 \end{tabular}\\[5pt]
  \begin{minipage}{8cm}
    \small Note: Reduced JD = JD - 2400000
  \end{minipage}

\end{table}

\section{Results and discussion}
\subsection { Light curves and frequency analysis}
The light curves of individual nights are shown in Figure 1 along with their corresponding
periodograms. A clear modulation at a period close to 26 minutes can be visually seen  in  all the light curves. Equivalently, analysis yields the presence of strong peak near 0.64 mHz in all the periodograms.
The mean $J$ magnitude of the program star, peak frequency and semi-amplitude of
the light curve on individual nights are presented in Table 2.
The peak frequency varies between 0.6367 mHz to 0.6439 mHz in periodograms of individual runs.
This variation of 0.007 mHz, however, is much below the error level estimated from half width at half maximum of the peak which is 0.02 mHz.

\begin{table}
 \centering
  \caption{Frequency analysis of daily lightcurves.}
  \medskip
  \begin{tabular}{@{}llll@{}}
   \hline
    Reduced JD of   & Mean J    & Peak       & Semi-      \\
    observation     & magnitude & frequency  & amplitude  \\
                   &           & (mHz)      & (mag)      \\
   \hline
	54452          & 15.54	   & 0.6367     & 0.14       \\
	54453          & 15.51	   & 0.6372     & 0.16       \\
	54483          & 15.53	   & 0.6413     & 0.11       \\
	54909          & 15.57	   & 0.6439     & 0.12       \\
	54910          & 15.51	   & 0.6418     & 0.12       \\
	55250          & 15.56	   & 0.6421     & 0.11       \\
   \hline

  \end{tabular}
\end{table}

Rapid but weak oscillations at frequencies other than that of the  principal component at 0.64 mHz were also observed in certain light curves. Notable among these is the one near 1.2 mHz which is the first harmonic of the spin period. In order to check the coherence of these weak oscillations we plotted the periodograms up to the Nyquist frequency for all the light curves of individual nights. We do not find any coherent frequency above 0.64 mHz. We thus conclude that these weak oscillations may be due to flickering in the light curve, a feature  which is a common characteristic of CVs.

\begin{figure}
\centerline{\includegraphics[width=12cm]{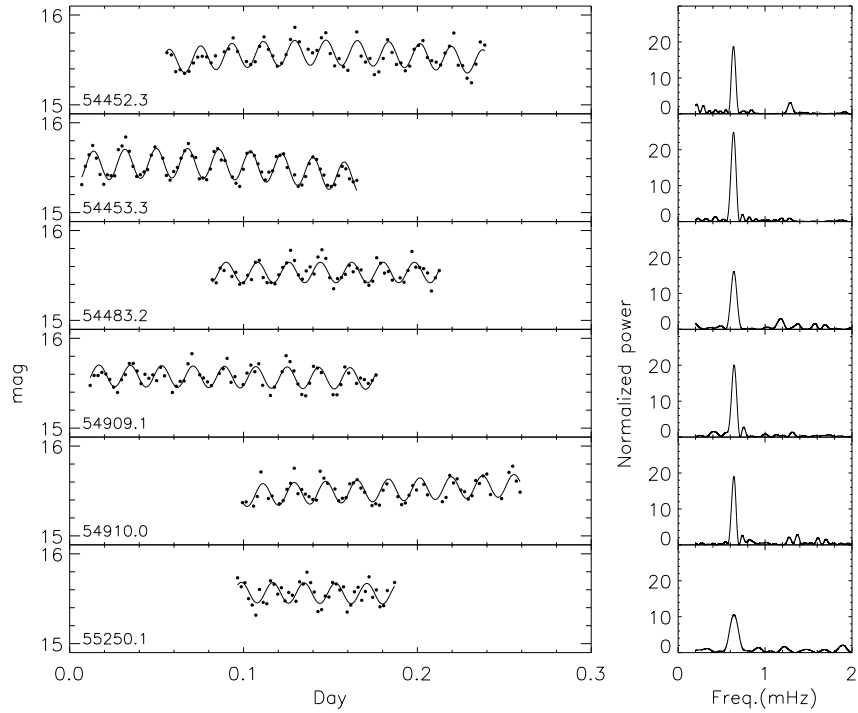}}
\caption{Light curves of all the nights listed in Table 1 are shown in the panels on the left.  The abscissa is HJD - HJD0 where HJD0 is arbitrary and given in the lower left for each day. Solid line shows the best fit sinusoid plus a polynomial fit representing the long term variation. The Lomb-Scargle periodograms of individual runs are shown in the right hand panels. It may be noted that the periodograms  extend up to the Nyquist frequency.\label{f:one}}
\end{figure}

The periodogram of the combined data plotted till the Nyquist frequency is shown in Figure 2. A strong peak can be seen near 0.64 mHz. Another noteworthy peak is at the low frequency end of the periodogram near 50 $\mu$Hz which is discussed subsequently.  One can also see minor peaks corresponding to 0.74 mHz and 1.28 mHz. But we consider  these latter peaks to be  statistically insignificant because the associated power is at a level of only 2 sigma above the background. Temporal gaps of various lengths in the observed data set cause the various aliasing peaks to be present in the periodogram of Figure 2. One day aliases are dominant in particular. This can be clearly seen in a plot of the window function which is presented as an inset in Figure 2. The periodogram between frequencies 0.4 and 0.9 mHz is shown in Figure 3 which is magnified part of the region containing peak near 0.64 mHz frequency. The highest power is at a frequency of 0.64134 $\pm$ 0.00008 mHz which corresponds to a period of 1559.2 $\pm$ 0.2 seconds. It may be noted that the formal error of the frequency (or period) is estimated by half-width at half-maximum of the highest peak in the peridogram which is the conventional way to determine the error.

\begin{figure}
\centerline{\includegraphics[width=12cm]{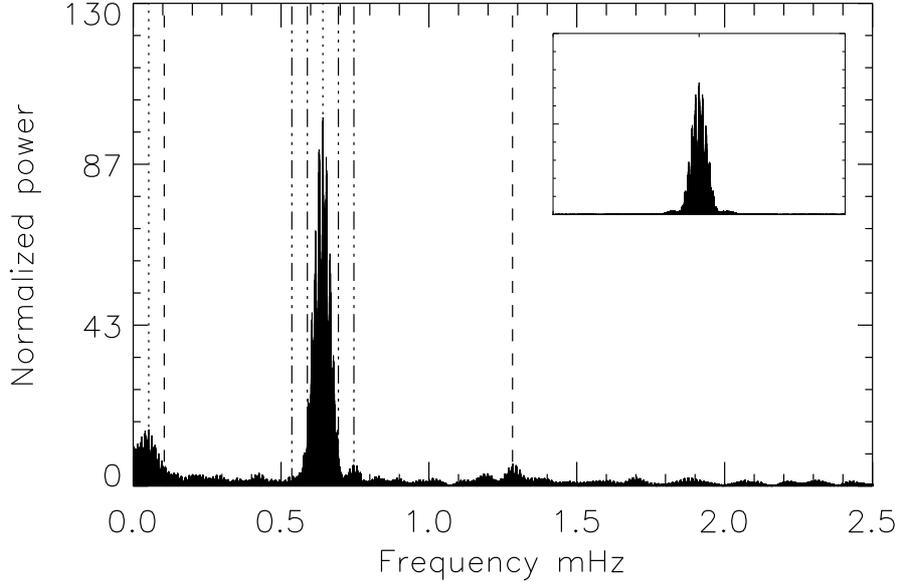}}
\caption{The Lomb-Scargle periodogram of all the data is shown. The dotted lines indicate orbital and spin
frequencies. The dashed lines show the first harmonic of the orbital and spin periods whereas the dot-dashed
lines represents the spin-orbital sidebands. The window function is plotted as an inset in the top of the figure.\label{f:two}}
\end{figure}

It is possible to generate the pulse profile of the spin modulation from the entire data set. As a first step, a best fit polynomial of either degree one or two was  subtracted from data of individual night's to remove the long term variation from the light curve. The data were then folded at a period 1559.2 second. The phase was divided into 25 bins to produce the spin pulse profile which is presented in Figure 4. The error in each bin is inferred from the standard deviation of the data points within the bin. From our analysis we find that the time of maximum of the oscillation is at RJD 54322.03408 $\pm$ 0.00005 d.

\begin{figure}
\centerline{\includegraphics[width=12cm]{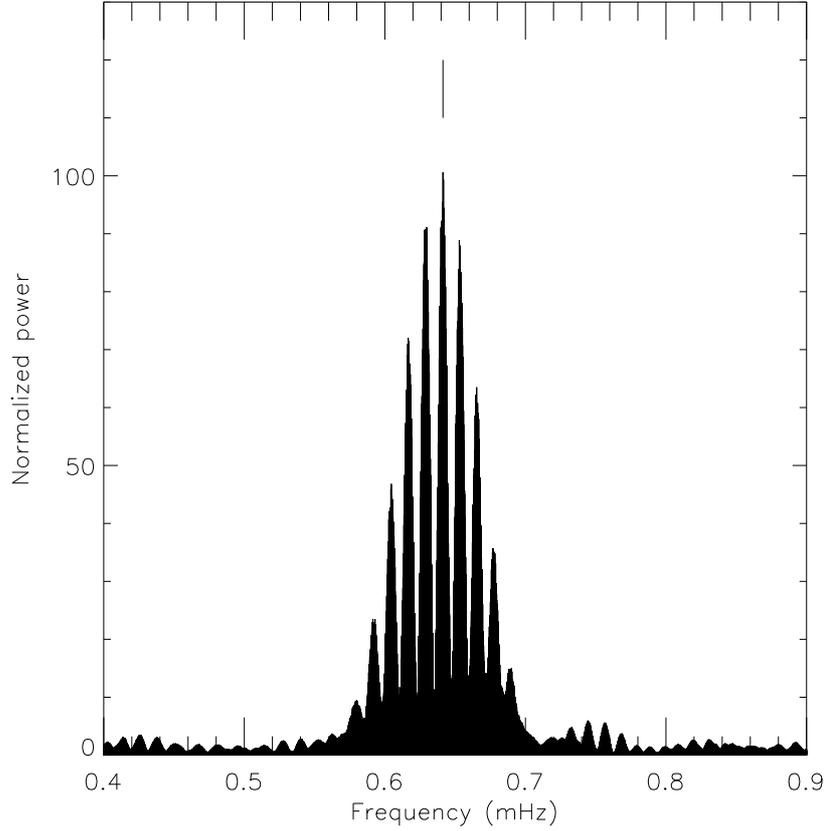}}
\caption{The LS periodogram of all the data near the spin frequency, which is marked with a vertical dash at 0.64134 mHz, is shown.\label{f:three}}
\end{figure}

\begin{figure}
\centerline{\includegraphics[width=12cm]{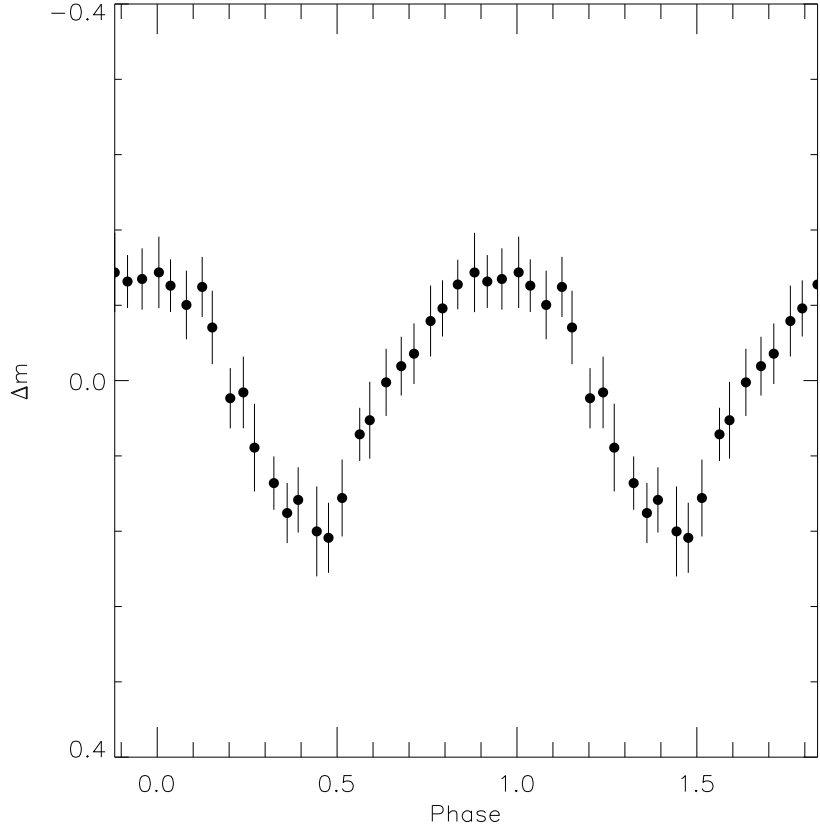}}
\caption{The binned pulse profile folded at 1559.2 seconds corresponding to the white dwarf spin period. The complete cycle is divided into 25 bins. The ordinate represents the relative magnitude and the abscissa
represents the fractional phase. The spin cycle is repeated once for clarity.\label{f:four}}
\end{figure}

In Figure 2, another significant peak is seen at a much lower frequency near 50 $\mu$Hz. A Lomb-Scargle periodogram  produced around this frequency, in the range 0.02 $\mu$Hz to 200 $\mu$Hz,  is shown in Figure 5. In this periodogram the maximum power is present at a frequency of 52.43075 $\mu$Hz
corresponding to a period of $\sim$5.3 hours. However, various alias peaks are strong in this spectrum and the 1-day alias in particular, at a frequency of 40.45529 $\mu$Hz or alternatively corresponding to a period of $\sim$6.9 hours,  is of comparable power to the 5.3 hour peak. It is thus possible that
either one of these periods (i.e 5.3 or 6.9 hours) may correspond to the orbital period but it is difficult to discriminate between the two based solely on our data.

However, we propose that the 5.3 h period more likely corresponds to  the orbital period of the binary system. First, the period of 5.3 hour is in reasonably good agreement with the orbital period of 5.54 h inferred by \citet{Schlegel05} from the separation of spin-orbit sideband peaks in the power spectrum of the X-ray data. Moreover a peak-like feature, albeit of weak power, is seen at a frequency 0.74478 mHz in Figure 2. If the 5.3 h period is correct, then such a feature is expected to arise as a result of the  $\omega$ + 2$\Omega$ sideband peak.  On the other hand, if the 6.9h period is correct, then based on the same argument advanced above, we should expect a feature at 0.73240 mHz which is however not seen. In summary, there are indications from our data  for a 5.3 h orbital period in the WX Pyx system.  The evidence for this is not overwhelming, but when results from the  X-ray data are also considered collectively, the 5.3 h period appears to be a realistic possibility. We thus assume that 5.30 $\pm$ 0.02 hour as the orbital period and adopt it, in following subsections, to determine some of the system parameters of WX Pyx. The orbital phase plot of WX Pyx is given in Figure 6 using a precise value of 5.297 hour(19072 second) for the orbital period. The time of the maximum light of the orbital modulation is found to be RJD 54327.1897 $\pm$0.0002 d.

\begin{figure}
\centerline{\includegraphics[width=12cm]{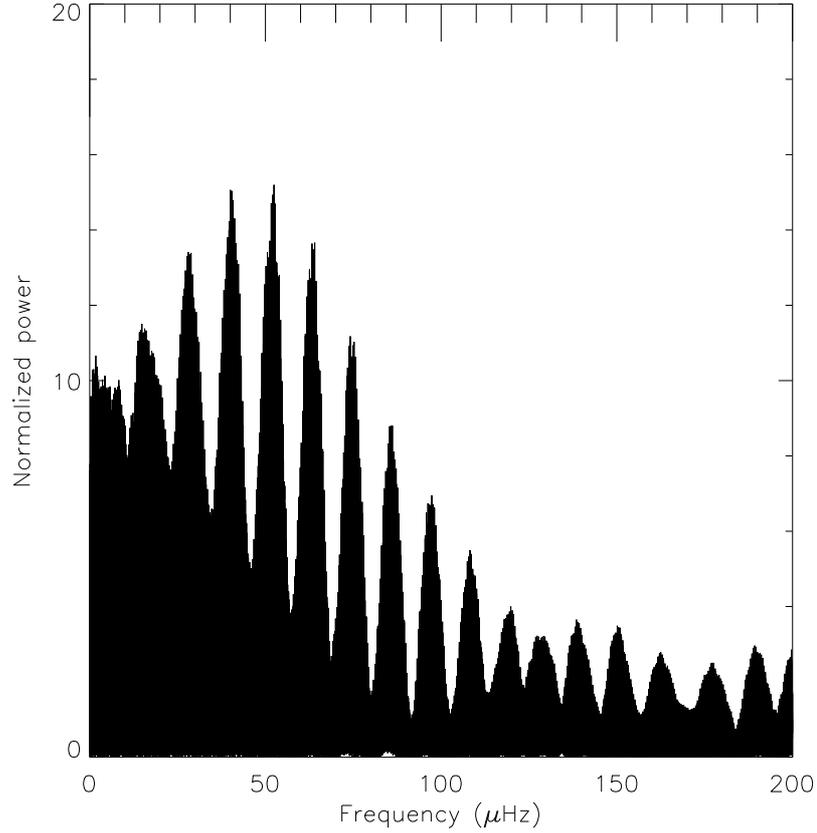}}
\caption{The LS periodogram near the orbital frequency is shown. It may be noted that there is no significant power above
the noise level at the first harmonic of the orbital frequency.\label{f:five}}
\end{figure}

\begin{figure}
\centerline{\includegraphics[width=12cm]{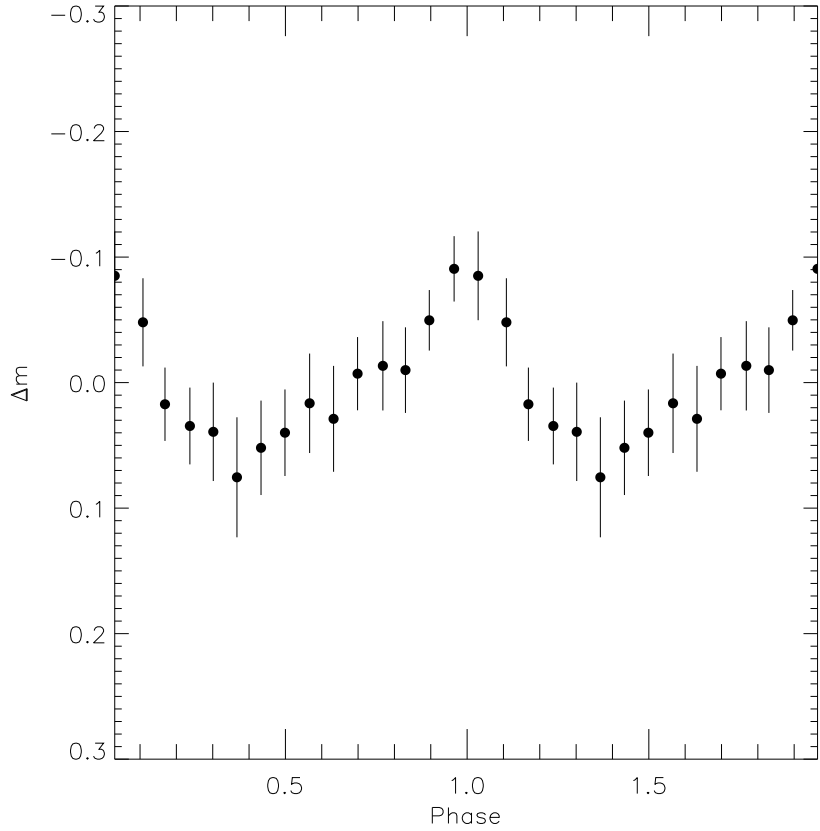}}
\caption{The binned phase plot of 5.30 hour corresponding to the orbital
period of WX Pyx is shown. The complete cycle is
divided into 15 bins. The ordinate represents relative magnitude and the abscissa represents the fractional
phase. The profile is repeated once for clarity.\label{f:six}}
\end{figure}

\subsection{Spectral type of the secondary}
It is well established that the secondary star in cataclysmic variables, with period less than 6
hours, resemble late-type main sequence stars i.e. stars of spectral type K or M. However, the
secondary stars of CVs have been shown to be a  little cooler than the isolated main sequence
star of same mass and radius (\citet{Knigge06} and Figure 7 therein; hereafter K2006). An empirical
relationship between orbital period and spectral type has been obtained (\citet{Beuermann98}; K2006).
Using the empirical relation discussed in K2006, and assuming an orbital  period of 5.3 hours,
we estimate the spectral type of the secondary to be constrained in the range  M2 $\pm$ 2.

\subsection{Distance to the source}
Since the spectral type of the system is known, one can estimate the distance to it using the
method first suggested by \citet{Bailey81}. This  inter-relates the surface brightness F$_\lambda$ at
wavelength $\lambda$, radius of the star $R$, distance $d$ and observed flux f$_\lambda$, by
following relation

\begin{equation}
    F_\lambda = f_\lambda \left({d \over R}\right)^2
\end{equation}

Assuming that the late type secondary star is the dominating contributor to the $K$  band flux
and expressing flux in magnitudes we can write,

\begin{equation}
    S_k = m_K + 5 - 5 log d + 5 log \left({R \over R_\odot}\right)
\end{equation}

where $m_K$ is the apparent magnitude and  $S_k$ is the surface brightness in the $K$ band . The radius $R$ of the star can be
calculated from the Roche-lobe geometry. However, the above approach gives only a lower limit
of the distance instead of distance itself because of our assumption that
the  secondary star is the sole contributor to the $K$  band flux. K2006 discussed the relations between
mass-radius and spectral type-orbital period and used these in the  above equation to derive the
simpler form given below viz.

\begin{equation}
    5 log (d_{lim}) = {m_K - M_K(P_{orb})+5}
\end{equation}

where $d_{lim}$ is the lower limit to the distance, $m_K$ is observed $K$ band magnitude and $M_{K}$
is the absolute magnitude as a function of orbital period.

From K2006 we obtain  an absolute $K$ band magnitude $M_K = 5.15$ for a 5.3 hour orbital period.
The  $K_s$ band magnitude of WX Pyx from 2MASS data ($K_s$ mag.  = 14.817) was converted to the CIT system
using the transformation given by \citet{Carpenter01} resulting in a transformed CIT $K$ magnitude of
14.857. The use of this value in the last equation  gives a lower limit to the distance as 874
pc. However, as pointed out by K2006, the assumption that the secondary contributes predominantly to
the $K$ band flux is not necessarily true in a general sense. K2006 studied the objects with
well-determined distances based on the parallax method and gave an offset value in magnitudes  which is to be applied to better estimate the distance to the object. Applying this offset of 1.22 magnitude for the $K$ band
we infer the distance to the source to be $\sim$ 1.53 kpc.

\subsection{Mode of accretion onto the white dwarf}
In their first optical follow-up observations, \citet{Hertz90} analyzed the optical spectrum
of WX Pyx and calculated the equivalent width of the H$\beta$ emission line. From this, they  estimated the $V$ band absolute magnitude of the disc M$_v$ using the relation given by \citet{Patterson84} which correlates these two parameters . They
compared M$_v$ with the observed $V$ band magnitude to infer that the  distance to WX Pyx  is $\sim$ 1.6 kpc. We keep in mind that  the relation given by \citet{Patterson84} holds for IPs with
well-developed discs  around the white dwarf. Interestingly the distance estimate in this work of $\sim$ 1.53 kpc, calculated using a different approach,  closely matches  that of
\citet{Hertz90}.  The agreement/consistency  of two independent methods yielding similar distance estimates to   WX Pyx suggests that the inherent assumption in the Hertz estimate (i.e. the system has a disc) is valid.  Thus it would appear that the WX Pyx  contains a partial disc around the white dwarf and it is mildly suggested that the mode of accretion is disc-fed. Further supporting evidence for the presence of the disc comes from the modulation at orbital period. The existence of such a modulation  can be explained only by  varying aspect of the bright spot formed at the
location of stream-disc interaction which, again, requires the presence of the disc.

\subsection{Constraints on the angle of inclination of the system}
It is expected that CVs show a ellipsoidal variation in their near-IR light curves at the first harmonic of the orbital period. Such a variation would arise due to the aspect variation of the Roche-lobe filled distorted secondary star. Fixing the other known parameters of the system one can infer the angle of inclination $i$ from the the shape of the phase curve. As seen in the Figure 1, the periodogram does not show significant power at the first harmonic of the orbital period shown as dashed line at the low frequency end of the spectrum. This suggests that the inclination angle of the binary system is low.

To estimate the upper limit of the inclination angle $i$ we produced a binned phase curve at the first harmonic of the orbital period. This phase curve shows a maximum scatter of $\pm$0.03 magnitudes. In order to estimate the upper limit of $i$ we produced the synthetic phase curve using Wilson-Devinney light curve code \citep{WD71}. Limb darkening values were obtained from interpolation of data in the table provided by \citet{Van93}. The mass ratio $q$ was varied between 0.2 and 0.6 and the appropriate inclination angle was obtained for each value of $q$ such that the synthetic light curve shows a similar scatter as that observed in the data. We find that for any value of $q$ in the chosen range,  the inclination angle $i$ is less then 30$^\circ$. This is suggestive that the inclination angle of the system is small.

\subsection{Mass transfer rate}
The mass transfer rate $\dot{M}$ of the system can be estimated by comparing the accretion
luminosity $L_{acc}$ and the released gravitational potential energy. When the mass is transferred
from the inner Lagrangian point to the white dwarf surface, the gravitational potential energy
is released and radiates away, largely at X-ray wavelengths in the case of magnetic CVs.

To estimate the X-ray luminosity of WX Pyx we used the flux values given by \citet{Evans07}. These authors modeled the X-ray spectrum of WX Pyx using  a combination of a non-thermal hard component and a thermal soft component. The unabsorbed bolometric X-ray flux corresponding to the soft and hard components are estimated to be $6.0\times10^{-13}$ ergs s$^{-1}$ cm$^{-2}$ and $7.51\times10^{-12}$ ergs s$^{-1}$ cm$^{-2}$ respectively giving a  total X-ray flux of $8.11\times10^{-12}$ ergs s$^{-1}$ cm$^{-2}$. This flux is converted into the X-ray
luminosity by the relation $L_x = 2{\pi}d^2F_x$ where $L_x$ is the X-ray luminosity in ergs s$^{-1}$, $F_x$ is the X-ray bolometric flux in ergs s$^{-1}$ cm$^{-2}$ and $d$ is the distance. Here, the factor $2{\pi}$ is used instead of $4{\pi}$ to take into account that the X-ray emitting shock region is close to the white dwarf surface and half of the total emission is blocked by the stellar surface. Using a value of $d$ = 1.530 kpc gives
the X-ray luminosity $L_x = 1.1\times10^{33}$ ergs s$^{-1}$. Estimation of $L_{acc}$ is a complex process requiring information about various aspects including the IR to UV emission from the accreting material,
absorption effect at UV and soft X-ray wavelengths and the contribution from X-rays above 10 keV \citep{Warner95}.
However \citet{Warner95} indicates that $L_{acc}$ may be estimated, within an uncertainty of a factor of 2,  by multiplying $L_x$
by  50. Therefore,  $L_x$ was multiplied by this factor  to estimate $L_{acc}$ which
in turn is
converted into the mass transfer rate $\dot{M}$ using  $L_{acc} = GM_{WD}\dot{M}/R_{WD}$ where
$L_{acc}$ is the total accretion luminosity and $M_{WD}$ and $R_{WD}$ are the mass and radius of the
white dwarf  respectively. $M_{WD}$ was chosen to lie in the range 1.31 to 1.4 $M_\odot$ as
inferred by \citet{Evans07} from modeling of the hard component of the X-ray data. $R_{WD}$
was calculated from the empirical mass radius relationship for white dwarfs given by
\citet{Pringle75}. We obtain the mass transfer rate to lie in the range 0.6 to 1.0$\times10^{17}$ g s$^{-1}$   or
 0.95 to 1.6$\times10^{-9} M_\odot$ yr$^{-1}$. This value of $\dot{M}$ is found to
be fairly typical for IPs as has been compiled for several other such objects \citep{Warner95}.

\subsection{Magnetic moment of the white dwarf}
Assuming a magnetic white dwarf with spherically symmetric accretion, the magnetospheric radius $r_{mag}$ is defined to be the radius at which the magnetic pressure balances the ram pressure of the infalling material. Following the calculation given in \citet{Frank02} the magnetospheric radius is given by

\begin{equation}
 r_{mag} = \left({{2{\pi}^2{\mu}^4} \over {\mu_0^2GM_{WD}{\dot{M}}^2}}\right)^{1/7}
\end{equation}

where $\mu$ and $\mu_0$ are magnetic moment of white dwarf and permeability of free space respectively. However, if the white dwarf is accreting via a disc, the true magnetospheric radius $R_{mag}$ can be estimated from $R_{mag}$ = 0.5 $r_{mag}$.

If the white dwarf of the IP is in spin equilibrium it is assumed that the magnetospheric radius $R_{mag}$ is very close to the co-rotation radius $R_{co}$ (e.g. Norton et al 2004). Here $R_{co}$ is the radius at which the accreting material in local Keplerian motion co-rotates with the magnetic field of the white dwarf. Therefore,
\begin{equation}
 R_{mag} \sim R_{co} = \left({{GM_{WD}P_{spin}^2} \over {4\pi^2}}\right)^{1/3}
\end{equation}
Comparing equations (4) and (5), we get
\begin{equation}
 \mu = 0.06 \left({{G^{5/6} M_{WD}^{5/6} P_{spin}^{7/6} \dot{M}^{1/2}}}\right)
\end{equation}

For WX Pyx, $\mu$ is thus estimated in the range $1.9$ to $2.4 \times 10^{33}$ G cm$^{3}$ for a $M_{WD}$ value ranging between  $1.31$ and $1.4 M_\odot$. This result is consistent with the magnetic moments of several other IPs estimated by \citet{Warner95}.

\citet{Norton08} have simulated the accretion flow of magnetic CVs in a detailed study. They have shown that for a fixed orbital period and mass ratio, different regions in a diagnostic plot of the spin period versus magnetic moment, correspond to different geometries for the accretion viz. disc, stream, ring and propeller (Fig 1 and 2 in \citet{Norton08}). For the orbital period and mass ratio specific for WX Pyx, we find  that the estimated value of magnetic moment along with the spin period falls well inside the region corresponding to a disc type accretion. This also supports disc-fed accretion as the favored mode of accretion in WX Pyx.

\section{Summary}
We have presented near infrared 1.25 $\mu$m $J$ band photometry of WX Pyx covering a total duration of 22 hours  spanning  six nights from December, 2007 to February, 2010. Our motivation was to determine the spin period of the object from IR observations, an exercise which has not been done earlier, and if possible to also determine the orbital period.  The frequency analysis of the light curve clearly indicates the presence of a spin period of 1559.2 $\pm$ 0.2 seconds for the white dwarf. However, the orbital period is less robustly determined. From the IR
observations, a  likely peak at 5.30 $\pm$ 0.02 hour is seen in the power spectrum of the object which is argued to be  the orbital period of the system. Subsequently, estimates are made of some of the physical properties and parameters of the system viz. the spectral type of the secondary star, the distance to the object, the mode of accretion, the angle of inclination of the system, the mass transfer rate and the magnetic moment of the white dwarf.

\section{Acknowledgments}
The research work at Physical Research Laboratory is funded by the Department of Space, Government of India.

\label{lastpage}


\begin{thebibliography}{}
\bibitem[\protect\citeauthoryear{Bailey}{1996}]{Bailey81} Bailey J., 1981, MNRAS, 197, 31
\bibitem[\protect\citeauthoryear{Bednarek $\&$ Pabich}{2010}]{Bednarek10} Bednarek W., Pabich J., 2011, MNRAS, 411, 1701
\bibitem[\protect\citeauthoryear{Beuermann}{1998}]{Beuermann98} Beuermann K., Baraffe I., Kolb U., Weichhold M., 1998, A$\&$A, 339, 518
\bibitem[\protect\citeauthoryear{Carpenter}{2001}]{Carpenter01} Carpenter J.M., 2001, AJ, 121, 2851
\bibitem[\protect\citeauthoryear{Das et al.}{2008}]{Das08} Das R.K., Banerjee D.P.K., Ashok N.M., Chesneau O., 2008, MNRAS, 391, 1874
\bibitem[\protect\citeauthoryear{Evans \& Hellier}{2007}]{Evans07} Evans P.A., Hellier C., 2007, ApJ, 663, 1277
\bibitem[\protect\citeauthoryear{Frank, King \& Raine}{2002}]{Frank02}Frank J., King A., Raine D.J., 2002, Accretion Power in Astrophysics Third ed., Cambridge University Press, Cambridge, p. 158
\bibitem[\protect\citeauthoryear{Hellier}{1994}]{Hellier94} Hellier C., 1994, PASP, 106, 209
\bibitem[\protect\citeauthoryear{Hertz \& Grindlay}{1984}]{Hertz84} Hertz P., Grindlay J.E., 1984, ApJ, 278, 137
\bibitem[\protect\citeauthoryear{Hertz et al.}{1990}]{Hertz90} Hertz P., Bailyn C.D., Grindlay J.E., Garcia M.R., Cohn H., Lugger P.M., 1990, ApJ, 364, 251
\bibitem[\protect\citeauthoryear{Hoard et al.}{2002}]{Hoard02} Hoard D.W., Wachter S., Clark L.L, Bowers T.P., 2002, ApJ, 565, 511
\bibitem[\protect\citeauthoryear{Knigge}{2006}]{Knigge06} Knigge C., 2006, MNRAS, 373, 484 (K2006)
\bibitem[\protect\citeauthoryear{Norton et al.}{2008}]{Norton08} Norton A.J., Butters O.W., Parker T.L., Wynn G.A., 2008, ApJ, 672, 524
\bibitem[\protect\citeauthoryear{O'Donoghue}{1996}]{ODonoghue96} O'Donoghue D., 1996, MNRAS, 278, 1075
\bibitem[\protect\citeauthoryear{Patterson}{1984}]{Patterson84} Patterson J., 1984, ApJS, 54, 443
\bibitem[\protect\citeauthoryear{Patterson}{1994}]{Patterson94} Patterson J., 1994, PASP, 106, 209
\bibitem[\protect\citeauthoryear{Pringle \& Webbink}{1975}]{Pringle75} Pringle J.E., Webbink R.F., 1975, MNRAS, 172, 493
\bibitem[\protect\citeauthoryear{Schlegel}{2005}]{Schlegel05} Schlegel E.M., 2005, MNRAS, 433, 635
\bibitem[\protect\citeauthoryear{Van Hamme}{1993}]{Van93} Van Hamme W., 1993, AJ, 106, 2096
\bibitem[\protect\citeauthoryear{Warner}{1995}]{Warner95} Warner B., 1995, Cataclysmic Variable Stars, Cambridge Astrophys. Ser., Cambridge University Press, Cambridge, p. 397
\bibitem[\protect\citeauthoryear{Wilson $\&$ Devinney}{1971}]{WD71} Wilson R.E., Devinney E.J., 1971, ApJ, 166, 605
\end{thebibliography}
\end{document}